\documentclass[aps,letterpaper,twocolumn,prl,footinbib,floatfix,showpacs,superscriptaddress]{revtex4-1}
\usepackage{amsmath}
\usepackage{amsfonts}
\usepackage{amssymb}
\usepackage[scanall]{psfrag}
\usepackage{psfrag}
\usepackage{graphicx}
\usepackage{color}
\usepackage[T1]{fontenc}
\usepackage{times}
\usepackage{ulem}
\usepackage[colorlinks,bookmarks=true,citecolor=blue,linkcolor=red,urlcolor=blue]{hyperref}

\begin{document}
\newcommand{\of}[1]{\left( #1 \right)}
\newcommand{\sqof}[1]{\left[ #1 \right]}
\newcommand{\abs}[1]{\left| #1 \right|}
\newcommand{\avg}[1]{\left< #1 \right>}
\newcommand{\cuof}[1]{\left \{ #1 \right \} }
\newcommand{\bra}[1]{\left < #1 \right | }
\newcommand{\ket}[1]{\left | #1 \right > }
\newcommand{\pil}{\frac{\pi}{L}}
\newcommand{\bx}{\mathbf{x}}
\newcommand{\by}{\mathbf{y}}
\newcommand{\bk}{\mathbf{k}}
\newcommand{\bp}{\mathbf{p}}
\newcommand{\bl}{\mathbf{l}}
\newcommand{\bq}{\mathbf{q}}
\newcommand{\bs}{\mathbf{s}}
\newcommand{\psibar}{\overline{\psi}}
\newcommand{\svec}{\overrightarrow{\sigma}}
\newcommand{\dvec}{\overrightarrow{\partial}}
\newcommand{\bA}{\mathbf{A}}
\newcommand{\bdelta}{\mathbf{\delta}}
\newcommand{\bK}{\mathbf{K}}
\newcommand{\bQ}{\mathbf{Q}}
\newcommand{\bG}{\mathbf{G}}
\newcommand{\bw}{\mathbf{w}}
\newcommand{\bL}{\mathbf{L}}
\newcommand{\ohat}{\widehat{O}}
\newcommand{\up}{\uparrow}
\newcommand{\down}{\downarrow}
\newcommand{\MM}{\mathcal{M}}
\author{Eliot Kapit}
\affiliation{Rudolf Peierls Center for Theoretical Physics, Oxford University, 1 Keble Rd, Oxford, OX1 3NP}
\author{Mohammad Hafezi}
\affiliation{Joint Quantum Institute, University of Maryland, College Park MD}
\author{Steven H. Simon}
\affiliation{Rudolf Peierls Center for Theoretical Physics, Oxford University, 1 Keble Rd, Oxford, OX1 3NP}

\title{Induced self-stabilization in fractional quantum Hall states of light}

\begin{abstract}
Recent progress in nanoscale quantum optics and superconducting qubits has made the creation of strongly correlated, and even topologically ordered, states of photons a real possibility. Many of these states are gapped and exhibit anyon excitations, which could be used for a robust form of quantum information processing. However, while numerous qubit array proposals exist to engineer the Hamiltonian for these systems, the question of how to stabilize the many-body ground state of these photonic quantum simulators against photon losses remains largely unanswered. We here propose a simple mechanism which achieves this goal for abelian and non-abelian fractional quantum Hall states of light. Our construction uses a uniform two-photon drive field to couple the qubits of the primary lattice with an auxiliary ``shadow" lattice, composed of qubits with a much faster loss rate than the qubits of the primary quantum simulator itself. This coupling causes hole states created by photon losses to be rapidly refilled, and the system's many-body gap prevents further photons from being added once the strongly correlated ground state is reached. The fractional quantum Hall state (with a small, transient population of quasihole excitations) is thus the most stable state of the system, and all other configurations will relax toward it over time. The physics described here could be implemented in a circuit QED architecture, and the device parameters needed for our scheme to succeed are in reach of current technology. We also propose a simple 6 qubit device, which could easily be built in the near future, that can act as a proof of principle for our scheme.
 
\end{abstract}

\maketitle

\section{Introduction}

The quantum simulation of exotic many-body states using photons is an important emerging field. By adding features such as 3- and 4-body interactions \cite{kempekitaev,ockoyoshida,kapitsimon,hafeziadhikari} or artificial gauge fields \cite{hafezidemler,nunnenkamp,kochhouck,umucalilarcarusotto,hafezilukin,kapitgauge}, the many-body ground states of these quantum simulators may become gapped and topologically ordered, with anyon excitations. However, these are open quantum systems, and will continuously leak photons into the environment, eventually reaching an empty vacuum state. To study the many-body physics of these systems, one must therefore devise a scheme to prepare the state and to refill hole states created by photon losses. In this work, we propose a simple mechanism which exploits the incompressibility of topological phases of matter to autonomously generate and protect non-equilibrium fractional quantum Hall (FQH) states of light. Our work is motivated by recent remarkable progress in circuit-QED systems \cite{Schoelkopf:2008p8712,Stajic:2013dh}, where the experimentally demonstrated strong nonlinearity between microwave photons can allow investigation of many-body effects \cite{Houck:2012iq}.

Following earlier work on dissipative state preparation \cite{diehlmicheli,krausbuchler} and ``dissipative gadgets" \cite{verstraete,pastawski}, we propose a generic construction which could stabilize the FQH ground states of strongly interacting lattice photons in an artificial gauge field. The engineered dissipation source in our case is an auxiliary ``shadow" lattice of qubits with a fast decay rate ($\Gamma_S$), with tuned energies and couplings so that hole excitations in the primary quantum simulator will be resonantly refilled. Since unwanted particle addition from thermal photons in the environment can be suppressed by simply lowering the system temperature, rapidly refilling lost photons is sufficient to protect the quantum many-body state. 

\begin{figure}
\includegraphics[width=.5\textwidth]{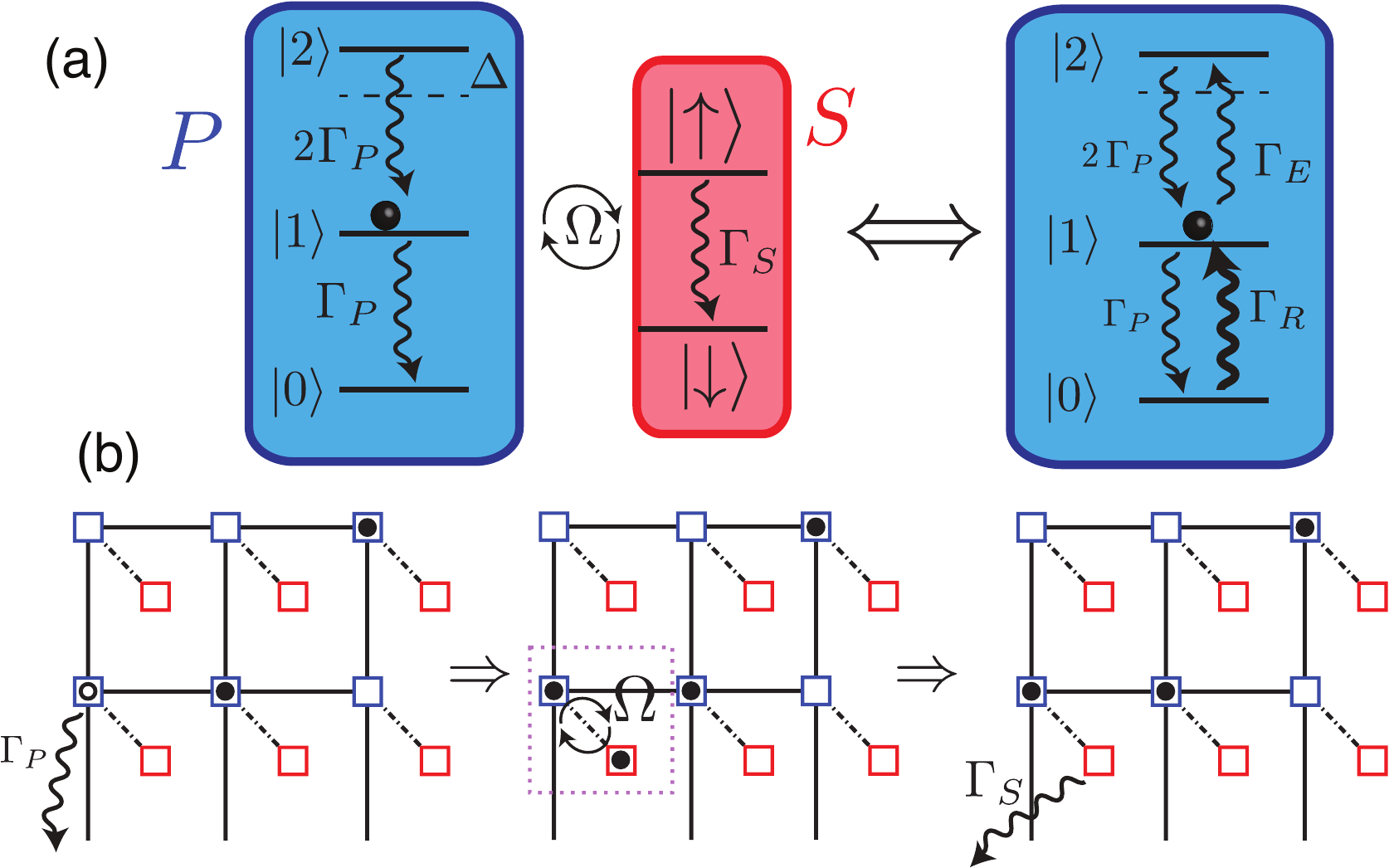}
\caption{(a) Simple example of dissipative protection of a quantum state. We consider a nonlinear resonator $P$, coupled to an auxilliary ``shadow" spin $S$, and we wish to hold the resonator in the one-photon state $\ket{1_P}$ against a loss rate $\Gamma_P$. We do so by coupling the resonator and shadow spin with a two-photon drive $\Omega$, which resonantly adds a photon to the resonator and excites the spin or removes a photon from the resonator and returns the spin to its ground state. The nonlinearity $\Delta \gg \Omega$ suppresses the $\ket{1} \to \ket{2}$ transition, and if the shadow spin has a fast relaxation rate $\Gamma_S \gg \Gamma_P$, then $\ket{1, \downarrow}$ is the most stable state of the system and all other configurations will rapidly relax toward it, as described in the introduction. (b) Many-body generalization of this mechanism to protect a lattice fractional quantum Hall system (blue) from losses. The primary quantum Hall lattice is coupled to a shadow lattice (red) through a 2-photon drive field, so that when a many-body hole excitation is created by a photon loss on the primary lattice (rate $\Gamma_P$), it is resonantly mixed with a \textit{particle} excitation on the shadow lattice (rate $\Omega$), which rapidly decays (rate $\Gamma_S$), returning the primary lattice to its many-body ground state. If $\Gamma_S \gg \Gamma_P$ and $\Omega \gg \Gamma_P$, holes will be refilled much more quickly than they are created, and the fractional quantum Hall ground state becomes the most stable state of the system. The many-body gap of the topological ground state suppresses further particle addition once the fractional quantum Hall state is reached.}\label{mainfig}
\end{figure}

The basic idea of our scheme is captured in FIG.~\ref{mainfig}a. To illustrate the dissipative protection of a quantum state, we consider a single nonlinear oscillator where $\omega_{12} = \omega_{01} + \Delta$. We wish to hold a single photon in the oscillator (state $\ket{1}$) against a loss rate $\Gamma_P$. We do so by coupling the oscillator to a single ``shadow" spin, and applying a parametric two-photon drive field, of strength $\Omega$, which resonantly adds a photon to the oscillator and excites the spin or removes a photon from the oscillator and returns the spin to its ground state. This coupling will rapidly add a photon to the oscillator whenever it is in state $\ket{0}$, but leaves state $\ket{1}$ unchanged; it cannot remove the photon since the rapid decay quickly returns the shadow spin to its ground state, and it cannot add a photon, since the nonlinearity brings the $\ket{1} \to \ket{2}$ transition far off resonance. Consequently, the effect of such coupling can be effectively described by a refilling rate $\Gamma_R$ and an error rate $\Gamma_E$ (corresponding to the weak $\ket{1} \to \ket{2}$ process), as shown in Fig.~1a. If we consider $\Delta \gg \Gamma_S \sim \Omega \gg \Gamma_P$, we have $\Gamma_E\ll\Gamma_P\ll \Gamma_R$. $\ket{1}$ is therefore the most stable state of the system with an average occupation probability $P_1 \sim 1-\Gamma_P/\Gamma_S$. This is in contrast to the conventional photon-blockade effect under a weak coherent drive, where the steady-state of the system is a superposition of different Fock states with a suppressed Poisson distribution \cite{Carmichael:2009tl}.

We will demonstrate both analytically and numerically that the many-body generalization of this simple mechanism can protect FQH states for arbitrarily long times, leaving only a small transient population of hole excitations due to the finite refilling rate. Similar to the simple system described above, in the FQH case it is the many-body gap which plays the role of the nonlinearity $\Delta$, ensuring that refilling stops when the FQH state is reached. Our method is particularly attractive for near-term experiments as it requires no state preparation sequence to initialize the system; since the ground state is protected by dissipation, one can simply turn on the shadow lattice coupling and wait for the system to relax into a topological ground state. We will also discuss possible extensions of the shadow lattice to protect other quantum simulator Hamiltonians, and demonstrate that the device parameters required for our scheme to succeed are within reach of current superconducting qubit technology.

\section{Fractional quantum Hall states of light}

The lattice fractional quantum Hall systems we consider are described by the combination of a topological flat (Chern) band \cite{kapitmueller,sungu,tangmei,neupertprl,rbprx,qi1,liuhighC,yaodipo,atakisioktel} and a local $k+1$-body interaction (where $k=1,2,3$) \cite{wangyao,wubernevig,bernevigregnault,liu,kapitbraiding}, and have topologically ordered ground states with anyon excitations. The total system Hamiltonian is given by
\begin{eqnarray}\label{FQH}
H_{P} &=& - \sum_{ij} J_{ij} \of{a_{iP}^{\dagger} a_{jP} e^{i \phi_{ij}}  + {\rm h.c.} }  \\ & &  + \frac{U_{k+1}}{\of{k+1}!} \sum_i \of{a_{iP}^{\dagger}}^{k+1} \of{a_{iP}}^{k+1}.  \nonumber
\end{eqnarray}
Here, the label $P$ denotes the primary lattice, and $a_{jP}^{\dagger}/a_{jP}$ creates or eliminates a photon at site $j$. If each site is a simple spin-$\frac{1}{2}$ degree of freedom, then the resulting hard core interaction yields an infinite $U_2$; higher order interactions (larger $k$) can be realized through more complex qubit arrangements \cite{kapitsimon,hafeziadhikari}. The hopping matrix elements are complex and model the Peierls phases of a uniform magnetic field which penetrates the plane; there are a number of proposals in the literature to engineer these phases \cite{hafezidemler,nunnenkamp,kochhouck,Umucallar:2012bo,Cho:2008ct,Otterbach:2010cp,Hayward:2012hg,umucalilarcarusotto,hafezilukin,kapitgauge}.These models are particularly robust if the Kapit-Mueller lattice Hamiltonian \cite{kapitmueller} (or a similar model devised by Ataki\c{s}i and Oktel \cite{atakisioktel}) is used to set $J_{ij}$, as in that case the band is exactly flat and spanned by lattice discretizations of the lowest Landau level (LLL) wavefunctions of the continuum. We will study the Kapit-Mueller Hamiltonian in the remainder of this work. We define the energy scale $\tilde{\mu}$ to be the combination of any chemical potential terms and zero point of the flat band dispersion, so that the energy of a particle in the LLL is $-\tilde{\mu}=O(J)$, where $\tilde{\mu}>0$ for the Hamiltonian $H_P$ as written in (\ref{FQH}). The magnetic length (which sets the characteristic size of the anyon excitations) is $l_B = 1/\sqrt{2 \pi \phi}$, where $\phi$ is the density of magnetic flux quanta per lattice plaquette.

We define the filling fraction $\nu = N/N_\phi$, where $N$ is the number of particles in the system and $N_\phi$ is the total number of flux quanta penetrating the lattice. We will assume that one of the $U_{k+1}$ is positive so that the interaction is repulsive, and set all the other $U$'s equal to zero. Whenever $N < N^* \equiv \frac{k}{2} N_\phi$, enough states remain in the LLL that particles can always be added at a minimum energy cost $- \tilde{\mu}$ (FIG.~\ref{mainfig}), but, ignoring edge effects, when $N= N^*$, the LLL is saturated, so to add an additional boson one must either allow more than $k$ particles to occupy a site or push bosons into the excited bands. In either case, this costs a finite amount of energy, so the system develops a many-body gap $\Delta$ when $N=N^*$. At this density, the bulk system's wavefunction is known exactly and given by the Read-Rezayi state of level $k$ \cite{readrezayi2}. We label this state $\ket{G_P}$, and for $k=1$ this state is nothing more than the $\nu=1/2$ bosonic analogue of the Laughlin state found in 2d electron gases; its excitations are quasiholes with half of the charge of the fundamental bosonic particles. For $k=2$ and $k=3$ the relevant states have non-abelian anyons (Ising and Fibonacci anyons, respectively), the collective states of which are topologically protected from local operations and can be used to encode and manipulate quantum information \cite{nayaksimon}.

\begin{figure}
\includegraphics[width=3.2in]{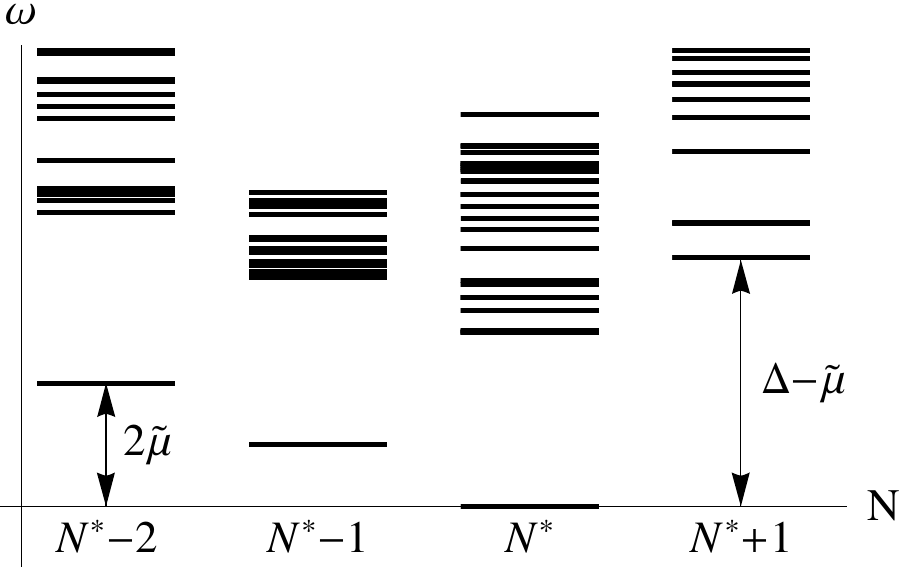}  
\caption{Generic level diagram of lowest Landau level bosons interacting through a repulsive contact interaction, governed by the Hamiltonian $H_P$ (\ref{FQH}). If the contact interaction is a $k+1$-body term, the system is gapless and degenerate until $N = N^* \equiv k N_\phi /2$, at which point the exact ground state is a Read-Rezayi state of level $k$ \cite{readrezayi2}. For all $N < N^*$, particles can always be added at a lowest Landau level energy of exactly $-\tilde{\mu}$, but when $N=N^*$ there is a large gap $\Delta$ (which scales as the smaller of the hopping $J$ and the interaction energy $U_{k+1}$) to any further particle addition. This property--that hole states are massively degenerate and isoenergetic-- reflects the fact that the anyonic quasiholes are dispersionless and noninteracting, and thus do not fractionalize on their own after being created locally through a single photon loss. Noninteracting topological defects are not a generic feature of most anyon systems, and the lack of interactions allows us to induce self-stabilization of the correlated states through the shadow lattice construction. }\label{levelfig}
\end{figure}

We would like to pause and point out the relative uniqueness of the spectrum shown in FIG.~\ref{levelfig}, which is a special bulk property of lowest Landau level bosons with a repulsive contact interaction. When a photon is lost, it creates a local group of anyonic fractional charges that sum to a whole boson, and if these fractional charges are pulled apart, they can no longer be eliminated through local perturbations. This is a generic property of anyonic systems, but unlike other systems such as Kitaev's honeycomb \cite{kitaev,knollekovrizhin} or FQH states with long-ranged interactions, the anyons in our model are noninteracting and dispersionless (they are composed entirely of LLL states for any set of anyon positions, and the parent wavefunction perfectly screens the interaction), and do not fractionalize on their own. The spectral density of hole excitations is thus a sharp band at $\omega = -\tilde{\mu}$ with contributions at all momenta, and as we will now demonstrate, this property allows for extremely efficient refilling.

\section{Photon losses and refilling}

Let us consider the effect of photon losses. We will assume that photons are lost at a uniform rate $\Gamma_P$ from each site in the lattice, independent of the energetics or configuration of the many-body state. The Hamiltonian (\ref{FQH}) conserves particle number, so there is no term to balance these losses, and all many-photon states will eventually relax to the empty state. To study FQH states, we must therefore introduce a new term which counteracts the losses. In order to highlight the necessity of our two-photon drive scheme,  we first consider the simplest mechanism for refilling, namely driving the system with a coherent drive: $ \mathcal{E} \cos (\omega_0 t) \sum_i \of{a_{iP}^{\dagger} + a_{iP} } $.  The eigenstates of the driven system will be resonant superpositions of many particle numbers, and will be largely unaffected by the losses. However, the weight of the strongly correlated state $\ket{G_P}$ within these superpositions decreases exponentially with increasing particle number $N$, so the signatures of FQH physics would only be visible in extremely small systems \cite{hafezilukin}. A simple coherent drive field is thus insufficient to prepare many-body FQH states of photons.

We now present an alternative construction which can stabilize $\ket{G_P}$ in arbitrarily large systems. We first introduce an auxiliary ``shadow" lattice of qubits, one for each site in the primary lattice, as shown in FIG.~\ref{mainfig}. Throughout this work, we will let the subscript $P$ designate operators acting on the primary lattice and $S$ designate operators on the shadow lattice.  The shadow lattice sites are uncoupled from each other, but coupled to the primary lattice sites through a parametric two-photon drive field of the form $ \Omega \cos 2\omega_0 t \sum_i \of{a_{iP}^{\dagger} \sigma_{iS}^{+} + a_{iP} \sigma_{iS}^{-}} $, where $\sigma_{iS}^{+(-)}$ adds (removes) a photon at shadow lattice site $i$.    Such parametric couplings can be achieved in any driven photonic system with $\chi^{(2)}$ and $\chi^{(3)}$ nonlinearities. For example, in circuit QED systems, such couplings have been experimentally demonstrated at a few-photon level, using Josephson parametric converters   \cite{Bergeal:2010iu,Anonymous:2013bq}.  We now let the rest frame excitation energies of sites in the primary and shadow lattices be $\omega_P$ and $\omega_S$. Passing to the rotating frame and discarding counter-rotating terms, the total system Hamiltonian becomes: 
\begin{eqnarray}\label{fullH}
H &=& H_{P} + H_{S} + H_{PS} + \of{\omega_P - \omega_{0} } \sum_i n_{iP}, \\ H_S &=& \frac{1}{2}\of{\omega_S -\omega_0 }\sum_{i} \sigma_{iS}^{z} \nonumber  \\
H_{PS} &=& \Omega \sum_i \of{a_{iP}^{\dagger} \sigma_{iS}^{+} + a_{iP} \sigma_{iS}^{-}}. \nonumber
\end{eqnarray}
Here, $n_{iP} = a_{iP}^{\dagger} a_{iP}$. When $N= N^{*} = k N_\Phi/2$, the primary lattice is in the FQH state $\ket{G_P}$, and there is a large gap $\Delta$ to all excitations created by photon addition ($a_{iP}^{\dagger}$). Similar to the discussion above, we define $-\tilde{\mu} = \omega_P - \omega_0 + O \of{J}$ to be the lowest Landau level energy in the rotating frame, i.e. the required energy to add/subtract a particle in the absence of interaction. The frequency choice of the rotating frame is arbitrary; however, we require that the two-photon resonant condition $\epsilon_S \equiv\omega_S - \omega_0 \simeq \tilde{\mu}$ holds. If $\Delta \gg \Omega$, then the combined state $\ket{G_P , 0_S}$ (where the primary lattice is in an FQH state and the shadow lattice is empty) is effectively unchanged by the drive field; photon addition is suppressed by the many-body gap, and photon removal by $H_{PS}$ is entirely forbidden since the shadow lattice is empty. Finally, we add the last ingredient: an engineered dissipation rate $\Gamma_S$ for every shadow lattice spin. We will now show that if $\Delta \gg \Omega \sim \Gamma_S \gg \Gamma_P$, the combined state $\ket{G_P , 0_S}$ is \textit{passively stable}, meaning that it will have the lowest decay rate of all states in the system's combined Hilbert space, and all other configurations will relax toward it over time.

Let us first consider the loss of a single photon in the primary lattice, creating a local pair of quasiholes near some site $j$ (for bosonic states in the Read-Rezayi sequence, the fundamental quasihole charge is $q/2$). As remarked earlier, this state $\ket{\Psi_{P}^{ \of{j}}}$ is an eigenstate of $H_P$, so the quasiholes will remain localized and not disperse on their own, and have an energy $+\tilde{\mu}$. However, when we turn on the two-photon drive ($\Omega \neq 0$), the quasihole pair leaves an open LLL state, so the two-photon drive can resonantly add a photon to it, refilling the hole and adding a photon to the shadow lattice as well (energy is conserved, as the shadow lattice energy is approximately $\tilde{\mu}$). The eigenstates of the combined system are thus resonant superpositions of a quasihole pair in the primary with an empty shadow lattice ($\ket{\Psi_{P}^{ \of{j}}, 0_S }$), and configurations with an FQH state in the primary lattice and a single photon at a shadow lattice site near $j$ ($\ket{G_P , 1_{jS}}$). Since the shadow lattice photon decays rapidly (rate $\Gamma_S$), the combined state has a much faster decay rate than $\ket{G_P , 0_S}$, and will relax back to it. Thus, the combined system has an induced decay rate for \textit{holes}, and if this rate is much faster than the hole creation rate $\Gamma_P$ then adding local holes will always increase the effective decay rate and relax the system back to $\ket{G_P , 0_S}$.

We would like to briefly remark on the issue of fractionalization in these systems. When a boson is lost from an FQH state, it leaves behind a pair of quasiholes, which are fractionally charged anyon excitations, for which refilling is exponentially suppressed if they are widely separated. However, the processes which would separate quasiholes in this system are higher order in $\Gamma_P/\Gamma_S$, so the creation rate for these configurations is very slow. Further, the population of individual anyons is self-limiting, since an increase in the anyon density increases the rate at which nearby pairs will be refilled by the shadow lattice. Thus, the equilibrium configuration of the system will be an FQH state with a small, transient population of quasiholes, as demonstrated below.

\begin{figure}
\includegraphics[width=3.5in]{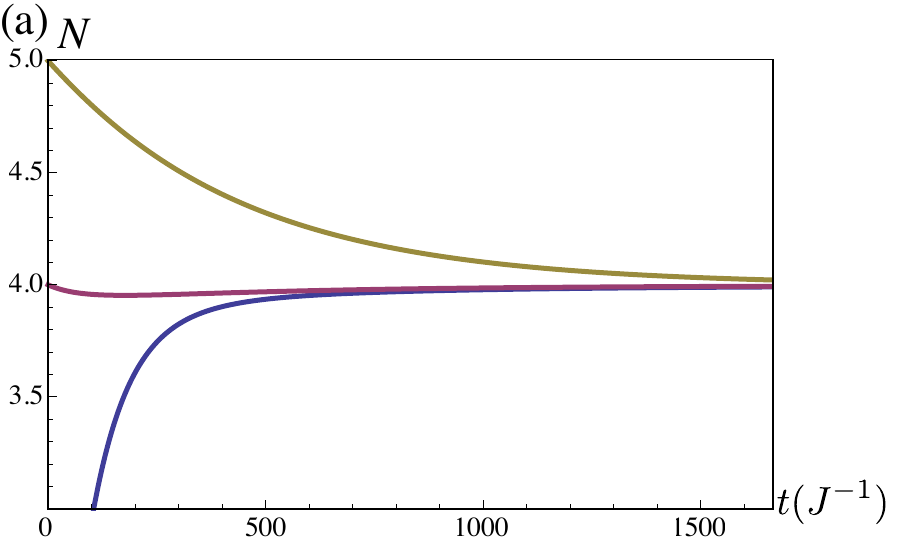}
\includegraphics[width=3.5in]{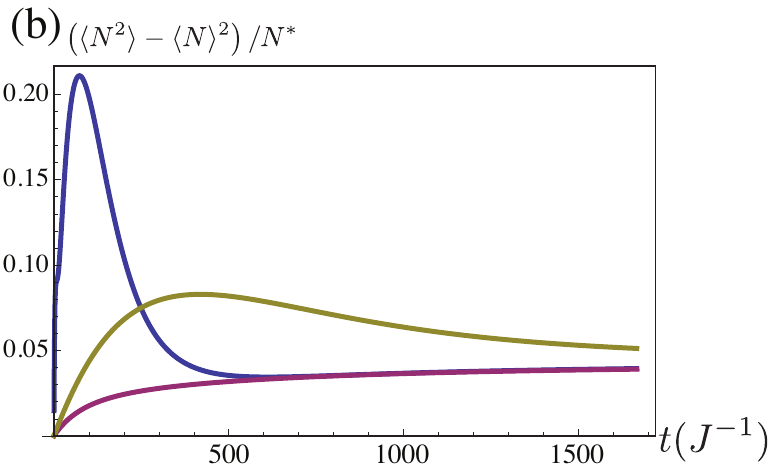}
\caption{(Color online) Demonstration that the Laughlin state is the most stable state of the system, computed in the refilling rate approximation for a $6\times4$ lattice with periodic boundary conditions, $\phi = 1/3$, $\Gamma_R/\Gamma_P \simeq 50$ and $\Gamma_P = 0.00067 J$. (a) The three curves initialize the system's density matrix in the empty state $N=0$ (blue, bottom curve), the Laughlin state (purple, center) and the Laughlin state with a quasiparticle excitation (gold, top), and time is evolved by numerically integrating the Lindblad equation (\ref{lindblad}). In all cases the system relaxes to a Laughlin state with a negligible density of excitations, as discussed in the text. Identical behavior was observed in all cases studied. (b) Signatures of incompressibility in this system. We plot the average number fluctuations $\of{\avg{N^2} - \avg{N}^2}/N^*$ for the same parameters and initial conditions, demonstrating that number fluctuations are highly suppressed by the combination of the energy gap $\Delta$ and the rapid refilling rate $\Gamma_R$. The small asymptotic value of $\of{\avg{N^2} - \avg{N}^2}/N^* \sim 0.05$ stands in contrast to the case of a gapless system subject to a coherent drive field, where $\of{\avg{N^2} - \avg{N}^2}/\avg{N}$ would be $O \of{1}$.}\label{stablefig}
\end{figure}

We can quantify the refilling rate $\Gamma_R$ with a simple Fermi's golden rule calculation. The complete elimination of the hole is a two step process (in which the first step, the transfer of the hole to a photon in the shadow lattice, is reversible) and limited by the slower of the two rates. For primary lattice states $\ket{\psi_{nP}}$ and $\ket{\psi_{mP}}$, these rates are:
\begin{eqnarray}\label{rates}
\Gamma_{ n \to m} &=& 2\pi \Omega^{2} \rho_{S} \of{\epsilon_{n} - \epsilon_{m} } \abs{\bra{\psi_{mP}} a_{jP}^{\dagger} \ket{\psi_{mP} } }^{2} \\
& & \times |\bra {1_{jS}} \sigma^+_{jS} \ket{0_{S}}|^2, \nonumber \\
\rho_S \of{\epsilon} &=& \frac{1}{\pi} \frac{\Gamma_S /2}{ \of{\epsilon - \epsilon_S }^{2} + \of{\Gamma_S / 2}^{2} }, \nonumber \\
\Gamma_{n\to m (tot)} &=& \frac{\Gamma_{n \to m} \times \Gamma_S}{\Gamma_{ n \to m}+ \Gamma_S}. \nonumber
\end{eqnarray}
For the refilling of a localized quasihole pair, the matrix element in $\Gamma_{n \to m}$ is typically $O \of{1}$, the shadow lattice matrix element is unity, and the energy is $\tilde{\mu}$, so the refilling rate $\Gamma_R$ reduces to
\begin{eqnarray}\label{refill}
\Gamma_R = \frac{4 \Omega^{2} \Gamma_S }{4 \of{\tilde{\mu} - \epsilon_S }^{2} + \Gamma_{S}^{2} + 4 \Omega^{2} }.
\end{eqnarray}
Analogous to the simpler case considered in the introduction and FIG.~\ref{mainfig}(a), if $\tilde{\mu} = \epsilon_S$ the process is resonant and refilling is extremely efficient. Conversely, if the energies do not match, the refilling rate is suppressed, preventing significant particle addition once the gapped state at $N=N^*$ is reached. For finite $\Omega/\Delta$, $\ket{ G_P , 0_S }$ will be weakly mixed with states with $N^* + p$ particles on the primary lattice and $p$ on the shadow lattice, and when a shadow lattice photon decays in one of these states it leaves a comparatively stable particle excitation in the primary lattice. The error rate $\Gamma_E$ of this process has the approximate form
\begin{eqnarray}\label{gammaE}
\Gamma_E = \frac{\Omega^{2} \Gamma_S}{\of{ \Delta + \tilde{\mu} - \epsilon_S}^{2}} + O \of{\Omega^{4}/\Delta^{4}}.
\end{eqnarray}
As $\tilde{\mu} \simeq \epsilon_S$, this reduces to $\Omega^{2} \Gamma_S / \Delta^{2}$, and we require it to be small, such that $\Gamma_{E} \ll \Gamma_P \ll \Gamma_{R}$ so $\ket{G_P , 0_S}$ remains passively stable. This error rate could be easily compensated through a slightly more complex shadow lattice construction \footnote{For example, we could add an additional transfer ($\pm$) coupling, of the form $g \sum_i \of{a_{iP}^{\dagger} \sigma_{iS}^{-} + a_{iP} \sigma_{iS}^{+}}$ between the primary and shadow lattice, and choose $\tilde{\mu} = \Delta/2$. In this case photons in the primary lattice can hop onto the shadow lattice (conserving the total number of photons), at an energy cost $\Delta$. Photons in an LLL state are thus far off resonant from the shadow lattice energy and will remain on the primary lattice, however, high-energy photons created by $\Gamma_E$ processes have energies near $\Delta$ and can thus be resonantly passed to the shadow lattice, where they will decay rapidly. The $\pm$ coupling needed to achieve this could be a simple capacitive interaction between the qubits of the two lattices in a circuit QED architecture, and would be easy to engineer.}, but in many cases $\Gamma_E$ is small enough that the primary loss rate $\Gamma_P$ can efficiently protect the system against incoherent particle addition. In all cases the maximum level of protection is set by the gap $\Delta$, even though the loss rate $\Gamma_P$ is independent of the many-body energetics of the primary lattice itself.

Taking all of these rates into account, the dynamics of the primary lattice are described by a  Lindblad equation \cite{gardinerzoller}. Considering the loss processes on both lattices and the Hamiltonian $H$ given by (\ref{fullH}), the system's density matrix $\rho$ evolves as
\begin{eqnarray}\label{lindbladfull}
\frac{\partial \rho}{\partial t} &=& -\frac{i}{\hbar} \sqof{H,\rho} + \frac{\Gamma_P}{2} \sum_j \of{2 a_{jP } \rho a_{jP }^{\dagger} - \cuof{ a_{jP }^{\dagger} a_{jP }, \rho} } \nonumber \\
& &+ \frac{\Gamma_S}{2} \sum_j \of{2 \sigma_{jS }^{-} \rho \sigma_{jS }^{+} - \cuof{ \sigma_{jS }^{+} \sigma_{jS }^{-}, \rho} }.
\end{eqnarray}
However, the full Hilbert space of the system is exponentially larger than that of the primary lattice alone, making (\ref{lindbladfull}) very expensive to evaluate in practice. We can construct a much more tractable set of equations by using the expressions (\ref{rates}-\ref{gammaE}) to integrate out the shadow lattice. In the Lindblad formalism, this amounts to adding a new set of modified quantum jump operators to capture the incoherent refilling process, with individual transition matrix elements appropriately rescaled to capture the energy dependence of the transition rates. In the basis of many body primary lattice eigenstates $\ket{\Psi_{nP}}$ we define the primary lattice quantum jump operators
\begin{eqnarray}\label{lindbladop}
\of{\tilde{a}_{j}^{\dagger}}_{nm} &\equiv & \frac{2\Omega}{\sqrt{\Gamma_{S}^{2} + 4 \Omega^{2} + 4 \of{\epsilon_{S} - \epsilon_{m} + \epsilon_{n}}^{2}}}  \of{a_{j P}^{\dagger}}_{nm}.
\end{eqnarray}
The primary lattice then evolves under the equation
\begin{eqnarray}\label{lindblad}
\frac{\partial \rho}{\partial t} &=& -\frac{i}{\hbar} \sqof{H_P,\rho} + \frac{\Gamma_P}{2} \sum_j \of{2 a_{j } \rho a_{j }^{\dagger} - \cuof{ a_{j }^{\dagger} a_{j }, \rho} } \nonumber  \\
& &+ \frac{\Gamma_S}{2} \sum_{j} \of{2 \tilde{a}_{j }^{\dagger} \rho \tilde{a}_{j } - \cuof{\tilde{a}_{j } \tilde{a}_{j }^{\dagger}, \rho} }. 
\end{eqnarray}
Using primary lattice eigenstates $\ket{\psi_{nP}}$ obtained analytically or through exact diagonalization, we can numerically integrate (\ref{lindblad}) to study the time evolution of the system due to the incohrent loss and refilling processes induced by decays on both lattices.

To confirm our analytical predictions, we conducted a series of numerical simulations, where we used exact diagonalization to generate the wavefunctions and system density matrix, which we then evolved with a Lindblad equation. We first considered both the full dynamics of the shadow lattice and the refilling rate approximation (\ref{lindbladop},\ref{lindblad}), and for the small systems studied ($4\times4$ lattices, due to the increased Hilbert space of the full Hamiltonian) the refilling rate approximation gives a quantitatively accurate description of the long-time and metastable equilibria. Final average densities typically agreed to within 2\% for $\Gamma_P \sim 10^{-3} \Delta$ and $\Omega \sim \Gamma_S \sim 25-100 \Gamma_P$. As the refilling rate approximation is far less computationally costly, we employed it in all our subsequent calculations. Likewise, in FIG.~\ref{stablefig}), we show that an FQH state with a small density of excitations is the equilibrium configuration of the system. We initialized the system (in this case, a $6 \times 4$ lattice with $N_\phi = 8$ and hard core two-body interactions) in three distinct configurations: the empty state with $N=0$, the Laughlin state at $N= N^* = 4$, and the Laughlin state with an extra quasiparticle excitation ($N=5$), and then let time evolve by numerically integrating (\ref{lindblad}). In all cases, the system relaxed to the same equilibrium configuration, though the approach to $N^*$ from the quasiparticle state was much slower than from the empty state, as $\Gamma_P = \Gamma_R/50$ for the parameters chosen.

\begin{figure}
\includegraphics[width=3.2in]{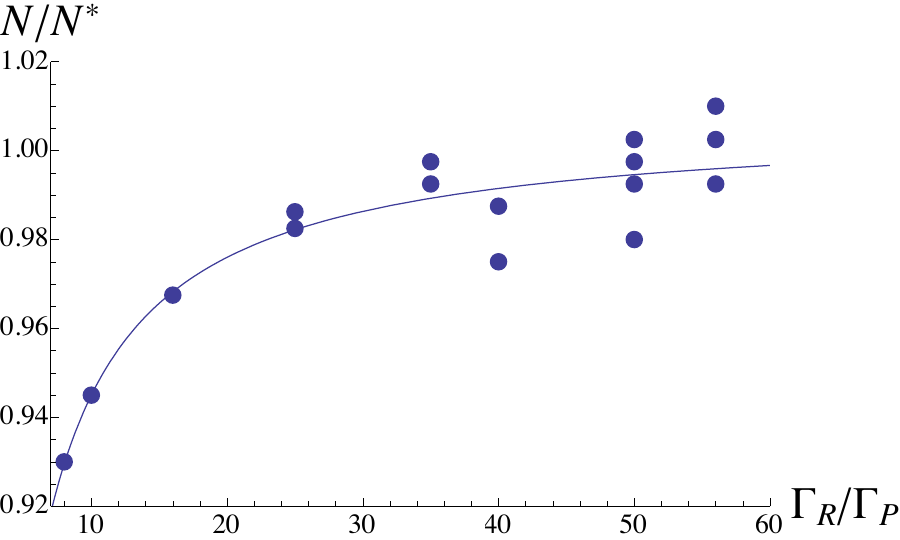}
\caption{Scaling of the equilibrium density vs. the ratio of the refilling rate $\Gamma_R$ to the primary loss rate $\Gamma_P$. The data (blue dots) are taken from simulations of $4 \times 4$, $5 \times 5$ and $6 \times 4$ lattices with $N_\phi =6$, 8 or 10 and hard-core two-body interactions, for various choices of $\Omega \sim \Gamma_S$ (with the refilling rate $\Gamma_R$ estimated from eq.~\ref{refill} with $\tilde{\mu}=\epsilon_S$), and $\Gamma_P$. Values of $N/N^* > 1$ are due to incoherent quasiparticle addition above the Laughlin state, which occurs at a rate $\Gamma_E \ll \Gamma_P$ as described in the text. The blue curve $ N/ N^* = 1.008 - 0.66 \Gamma_P / \Gamma_R$ is a numerical fit to the data, and the coefficient of 0.66, rather than a number close to unity, stems from geometric considerations (the finite size of quasiholes means that multiple sites contribute to their refilling, while a single site contributes to their creation for a given loss event) and the contribution from incoherent quasiparticle addition rates. The incoherent addition rate increases with increasing $\Omega$ and $\Gamma_S$, and thus generically grows with increasing $\Gamma_R$ for a given $\Gamma_P$.}\label{fig:scaling}
\end{figure}

In the long-time limit, the density matrix $\rho$ of this system satisfies detailed balance, and can be well-approximated by a thermal ensemble. Specifically, if $\Gamma_{R}/\Gamma_P \gg 1$ and $\Gamma_P/\Gamma_E \gg 1$, the detailed balance condition dictates that the probability $P_{N}$ of finding $N$ particles in the system will be suppressed by a factor of $\of{\Gamma_P/\Gamma_R}^{N^* - N}$ (or $\of{\Gamma_E/\Gamma_P}^{N- N^*}$ for $N > N^*$) relative to the probability $P_{N^*}$ of finding the system in a fractional quantum Hall state. Average final densities as a function of $\Gamma_R/\Gamma_P$ are plotted in FIG.~\ref{fig:scaling}, demonstrating that the density of quasihole excitations scales as $\Gamma_P/\Gamma_R$ times a constant which depends on the flux density $\phi$.

\begin{figure}
\includegraphics[width=3.2in]{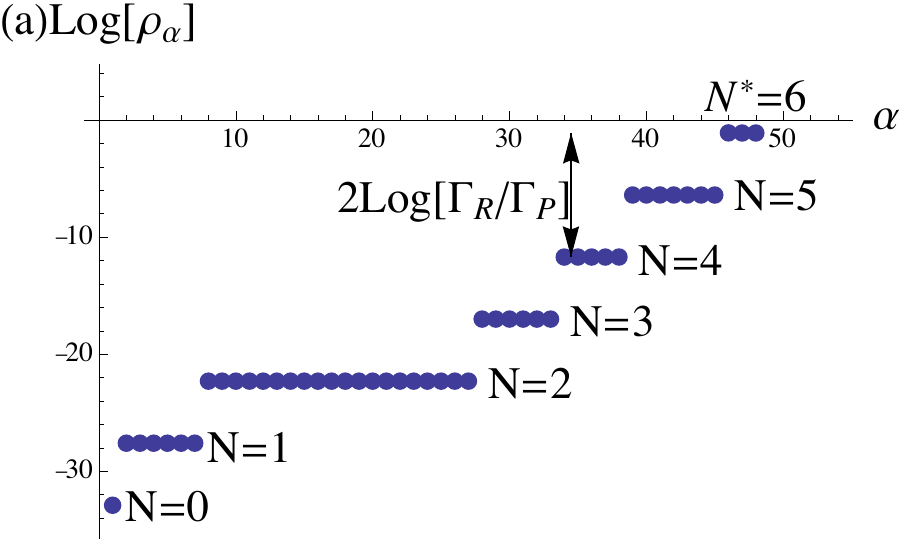}
\includegraphics[width=3.2in]{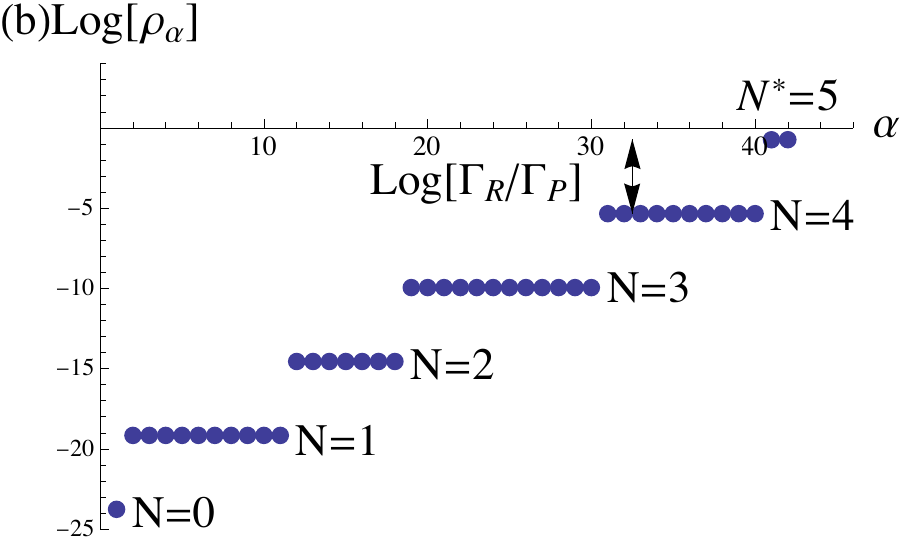}
\caption{Long-time stabilization of (a) Moore-Read and (b) Laughlin states. (a) Using the refilling rate approximation (\ref{lindbladop},\ref{lindblad}), we studied the dynamics of a $6 \times 4$ lattice with periodic boundary conditions, $\phi = 1/4$ and a hard-core three-body interaction. In this limit, the exact ground state at $N=N^*=6$ is a Moore-Read state \cite{mooreread}, which is threefold degenerate on the torus. In the figure, we plot the logarithm of the ordered eigenvalues $\rho_\alpha$ describing the equilibrium occupation of the lowest Landau level states (other states not shown) for $\Gamma_R = 200 \Gamma_P$ and $\Gamma_R \ll \Delta$. Each level corresponds to a fixed particle number $N$, with $N=6$ being the Moore-Read state. The equilibrium occupations of the LLL states obey detailed balance and can be modeled by an induced chemical potential and temperature $\mu_{ind}/T_{ind} \simeq \log \of{\Gamma_{R}/\Gamma_P }$. The final occupation of the three Moore-Read states was 99\% in this case. In such a small system, equilibration was rapid, but as remarked in the text, the exponential suppression of fractionalization in large systems leaves the equilibration rate an open question. (b) Long-time configuration of a $5 \times 5$ lattice with $N_\phi = 10$ and hard-core 2-body interactions, with $\Gamma_R = 100 \Gamma_P$. The occupation of the two Laughlin states at $N = N^* = 5$ is 95\%.}\label{pfaffianfig}
\end{figure}

Likewise, we can model the equilibrium density matrix $\rho$ as a thermal density matrix with an induced (positive) chemical potential $\mu_{ind}$ (replacing the LLL energy $\tilde{\mu}$) and temperature $T_{ind}$, as shown in FIG.~\ref{pfaffianfig}. The positive chemical potential causes quasihole pairs to be gapped, with the ratio $\mu_{ind}/T_{ind} \simeq \log \of{\Gamma_{R}/\Gamma_P }$ (and $\of{\Delta-\mu_{ind}}/T_{ind} \simeq \log \of{\Gamma_P/\Gamma_E}$). The average occupations of all LLL states at a given $N$ are equal, and we observed this behavior in models with Laughlin ($U_2 \to \infty$) and Pfaffian ($U_3 \to \infty, U_2 = 0$) ground states of up to eight particles on small lattices of up to 25 sites with periodic boundary conditions, computed using the refilling rate approximation. Example density matrices for equilibrium configurations of these lattices are shown in FIG.~\ref{pfaffianfig}.

It is important to note that in the presence of impurities (which locally prevent refilling and bind anyons) or nontrivial boundary conditions, these systems have topologically degenerate ground states. These states are mixed by losses and refilling, but this is a weak process, since the fractionalization of quasiholes is exponentially suppressed, particularly at high flux densities. In the small systems which were accessible to us through exact diagonalization, the exponential tails of the relevant correlation functions are long enough to cause significant mixing through single-boson operations, but in large systems this should no longer be the case. The degree to which the shadow lattice could passively protect quantum information encoded in the topological ground state degeneracy of large systems (and the degree to which these systems thermalize, given the relative inaccessibility of highly fractionalized configurations) is thus an open question. A future study which used the exact solubility of the Kapit-Mueller Hamiltonian to generate the eigenbasis analytically (likely evaluating the wavefunctions themselves through Monte Carlo methods) could  probe fractionalization effects more directly in much larger systems, but is beyond the scope of this work.

\section{Implementation through circuit QED}

As remarked earlier, we feel that circuit QED architectures provide a promising path to realizing the system described in this article. Using modern charge-insensitive qubit designs such as flux, transmon or fluxonium qubits \cite{Schoelkopf:2008p8712,Stajic:2013dh} to construct the primary lattice, the two-photon drive field could be readily implemented through a set of Josephson parametric amplifiers \cite{Bergeal:2010iu,Anonymous:2013bq}, and the artificial gauge field could be engineered through appropriate patterns of phase shifted drive fields \cite{kapitgauge,hafeziadhikari}. The next nearest neighbor couplings required to implement the Kapit-Mueller Hamiltonian could be realized through a multilayer fabrication process. Periodic boundary conditions could also be engineered in this manner, though they would be technically challenging to implement. Given the excitation energies and nonlinearities of these qubits, nearest neighbor couplings $J/h$ up to $2\pi \times 100$MHz are feasible, and primary lattice decay rates $100 {\rm kHz} \geq \Gamma_P \geq 10 {\rm kHz}$ are readily achieved in planar qubit architectures. This in turn suggests an upper limit of $\Delta/\Gamma_P \sim 10^4$, allowing for extremely effective refilling.

The most straightforward experimental probes to demonstrate the presence of an FQH state are density measurements. As the system is gapped at $N^* = k N_\phi /2$ (where $N_\phi$ is the number of flux quanta in the system and the system Hamiltonian exhibits $k+1$-body contact interactions), the small number fluctuations $\avg{N^2} - \avg{N}^{2} = O \of{\Gamma_{P}^2/\Gamma_{R}^2}$ at $N \simeq N^*$ would be a clear signature of strong correlations. To confirm that the gapped state is an FQH state, one could vary the flux density $\phi$ of the artificial gauge field. As the FQH state occurs at $N^* = k N_\phi /2$, the equilibrium density of the system will track the flux density as it is increased or decreased. We caution, however, that obtaining the flat band in the Kapit-Mueller Hamiltonian requires the magnitudes of the next nearest neighbor couplings to depend on the flux density, but for small changes in $\phi$ the resulting small bandwidth should not disrupt the FQH physics. Ranged density correlation functions could likewise be used to shed light on the underlying topological state.

\section{A simple device to demonstrate passive error correction}

Finally, while the quantum Hall systems we have considered would be tremendously exciting to realize, and the device parameters required for the shadow lattice to work have already been attained in previous experiments, the sheer size and complexity of these circuits present real challenges. It is thus worth considering models which could demonstrate the fundamental result of this work-- that a properly-tuned single particle shadow lattice can protect interesting many-body states against photon losses-- using only a handful of qubits. We now present one such implementation. 

\begin{figure}
\includegraphics[width=3.0in]{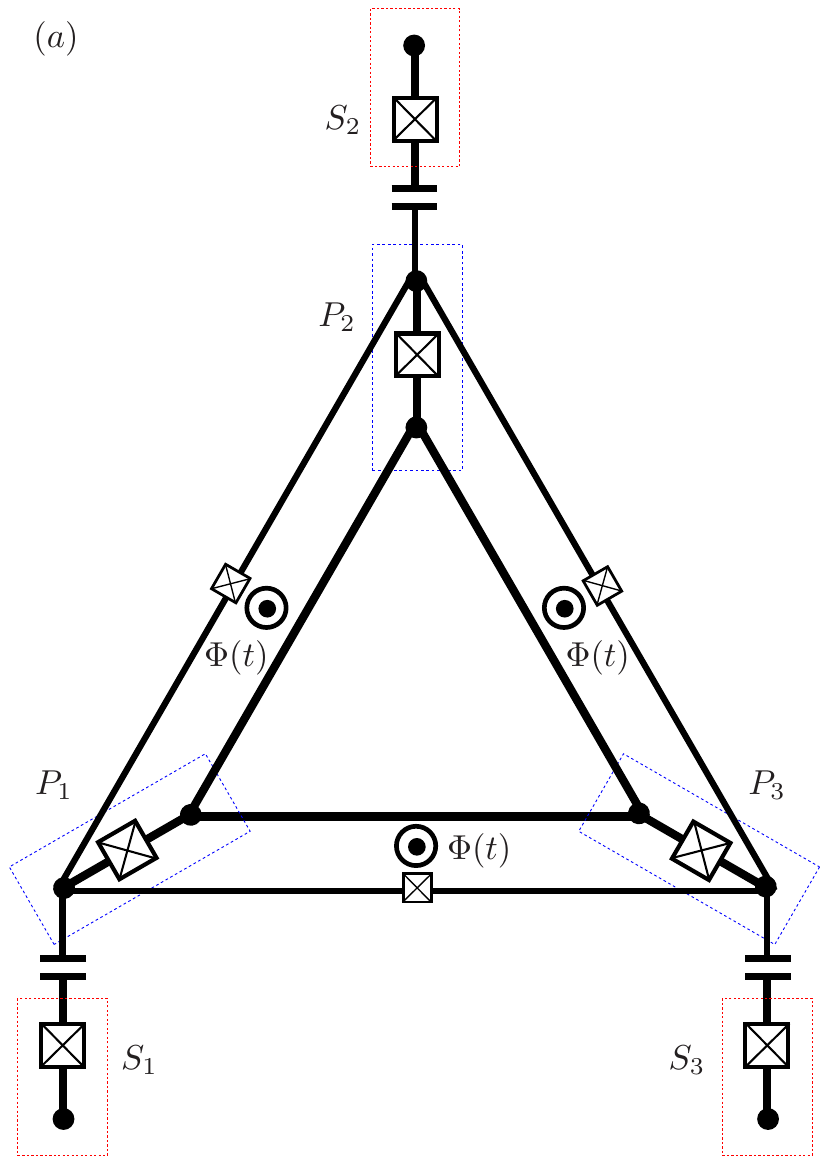}
\includegraphics[width=3.0in]{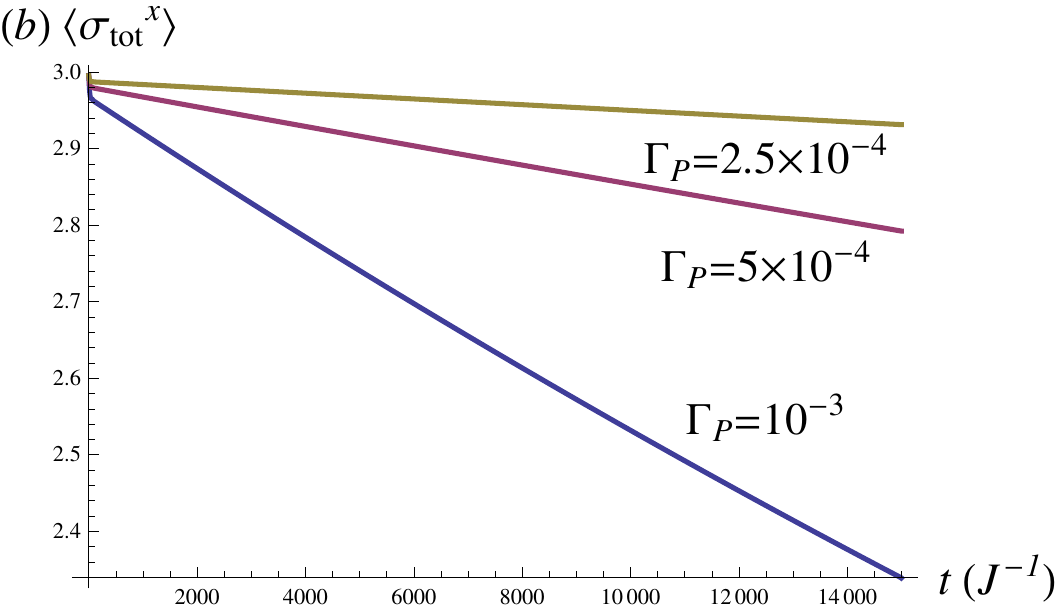}
\caption{(Color online) A simple circuit to demonstrate passive error correction. (a) Ring of three primary superconducting transmon qubits ($P_1,P_2,P_3$, blue), coupled capacitively to three shadow qubits ($S_1,S_2,S_3$, red). The small Josephson couplings which link the primary qubits are biased by a time-dependent flux $\Phi \of{t}$, which implements the two-photon drive field. Combined with the internal capacitances of the junctions, the resulting interaction is $- J \sigma_{iP}^{x} \sigma_{jP}^{x}$ in the rotating frame. As described in the text, the shadow lattice qubits rapidly correct single bit flip errors, protecting both of the degenerate ground states of the primary lattice Hamiltonian. (b) Demonstration of state protection for this circuit, with $g = 0.05 J$ and $\Gamma_S = 0.1 J$ ($\Gamma_R = 0.05 J$). We initialized the system in the ground state $\ket{111}$ and numerically integrated the Lindblad equations for the full system's evolution. For $\Gamma_P = \cuof{10^{-3}, 5 \times 10^{-4}, 2.5 \times 10^{-4}} J$, the observed decay rates for the ground states $\ket{000}$ and $\ket{111}$ were $\cuof{4.6 \times 10^{-5}, 1.3 \times 10^{-5}, 3.8 \times 10^{-6}}$, demonstrating the effectiveness of the error correction. }\label{ringfig}
\end{figure}

We consider a primary ring of three superconducting qubits, as shown in FIG.~\ref{ringfig}(a). We couple each of these qubits with a parallel combination of the two-photon drive field described earlier and an ordinary capacitive coupling, which takes the form $\sigma_{iP}^{+} \sigma_{jP}^{-} + \sigma_{iP}^{-} \sigma_{jP}^{+}$ in the rotating frame. This coupling could be straightforwardly achieved through the flux biased Josephson junction coupling shown in FIG.~\ref{ringfig}(a). If we choose $\Phi \of{t} = \frac{\Phi_0}{4} \of{1 + f \cos 2 \omega t}$, where $\Phi_0$ is the superconducting flux quantum, $f \ll 1$ and $\omega$ is the excitation energy of the qubits, the only part of the Josephson coupling which survives in the rotating frame is the two-photon drive term. We choose the coefficients of both the two-photon and exchange terms to be equal, and their resulting sum is a pure $\sigma_{iP}^{x} \sigma_{jP}^{x}$ interaction with energy $-J$. We then add three additional shadow qubits (which could be simple resonators in this case), which are coupled through a weak capacitive interaction with their corresponding primary qubits, and choose their excitation energies to be $4J$ higher than the energies of the primary qubits. Our total rotating frame Hamiltonian is:
\begin{eqnarray}\label{Hring}
H &=& - J \of{\sigma_{1P}^{x} \sigma_{2P}^{x} + \sigma_{2P}^{x} \sigma_{3P}^{x} + \sigma_{3P}^{x} \sigma_{1P}^{x}} \\
& & + \sum_{i=1}^{3} \sqof{ g \of{ \sigma_{iP}^{x} \sigma_{iS}^{x} + \sigma_{iP}^{y} \sigma_{iS}^{y}} + 2 J \sigma_{iS}^{z} }. \nonumber
\end{eqnarray}

We now let $J \gg g \simeq \Gamma_S \gg \Gamma_P$, replicating the scale hierarchy of the many-body shadow lattice. The ground state manifold of (\ref{Hring}) is 2-fold degenerate, and up to tiny corrections from the coupling to the shadow lattice, the ground states are simply $\ket{000}$ and $\ket{111}$ in the $\sigma^{x}$ basis, in both cases with all shadow qubits in their ground states. When a photon is lost from the primary lattice, the resulting $\sigma_{iP}^{-} = \sigma_{iP}^{x} - i \sigma_{iP}^{y}$ operation can tip the system into one of its excited states, but thanks to the $\sigma_{iP}^{y} \sigma_{iS}^{y}$ coupling to the shadow lattice, the true excited states of the system are resonant superpositions of a spin flip on the primary lattice with the corresponding shadow qubit in its ground state, and the primary lattice in its ground state with an excitation in one of the shadow qubits. As in the quantum Hall lattice, the shadow qubits rapidly relax, protecting the primary lattice ground state manifold through engineered dissipation.

This circuit thus replicates the ``textbook" three-qubit bit flip code, with the energy selectivity of the shadow lattice taking over the role of an observer acting in response to measurements. If the refilling rate $\Gamma_R$ is fast compared to $\Gamma_P$, then mixing between $\ket{000}$ and $\ket{111}$ (again, in the $x$ basis) will be heavily suppressed, as a second spin flip must occur before the first flip is corrected in order to induce a transition between the two ground states. If we thus take $\ket{000}$ and $\ket{111}$ as our logical states, the shadow lattice is a genuine source of quantum error correction, protecting the system against bit flip errors. This protection is demonstrated in FIG.~\ref{ringfig}(b); beginning in the $\ket{111}$ state, we integrated the Lindblad equations for the full six-qubit system to show that the shadow lattice suppresses the rate at which the two spin states are mixed by a factor of $\Gamma_P/\Gamma_R$. As $\Gamma_R > 10$MHz is achievable in this setup, it is potentially much more effective than measurement-based approaches.

However, like the three-qubit bit flip code, it does \textit{not} protect the logical qubit against dephasing. For an arbitrary superposition $\ket{\psi} = \alpha \ket{000} + \beta \ket{111}$, the magnitudes $\abs{\alpha}$ and $\abs{\beta}$ are protected by the shadow lattice, but the relative phase $\arg \of{\alpha/\beta}$ is not; the $\sigma_{iP}^{x}$ component of a photon loss operation does not excite the primary qubits, but it returns opposite signs for the two ground states, dephasing the superposition. More complex constructions could be introduced to protect the system from dephasing as well, but they would require higher order $n$-qubit interactions (where $n>2$) and are thus much more difficult to engineer.
 
\section{Discussion and Extension to Other Systems}

In this work, we have presented a simple construction which can protect a topologically ordered, anyonic state of photons against losses without any active intervention from external observers. While bosonic fractional quantum Hall states are particularly exotic and fascinating phenomena, they are by no means the only system which could be stabilized through a shadow lattice construction. The shadow lattice is fundamentally a local energy pump, which is tuned to be extremely efficient for a narrow energy range. If photon losses or decoherence create finite energy excitations in a primary lattice, and the coupling to the shadow lattice acts as the inverse of the error process, then these excitations can be rapidly eliminated so long as the energies on the two lattice match. Due to the unique spectrum of lowest Landau level bosons with contact interactions, a single shadow lattice energy is enough to stabilize the state. 

Other primary lattice Hamiltonians could similarly be passively stabilized, provided that they have a many-body gap $\Delta$. Models where the hole excitations are dispersive, for example, could be stabilized by adding dispersion to the shadow lattice through hopping terms between shadow lattice sites; if the many-body dispersion of holes in the primary lattice roughly matches the single-particle dispersion on the shadow lattice, refilling will still be very efficient. However, Hamiltonians where defects are strongly interacting would be much more difficult to stabilize than the FQH system considered here. A 1d Ising chain of more than three sites is the simplest model which falls into this class: single spin flips could be eliminated rapidly, but two or more adjacent spin flips produce extended domains that a single-band shadow lattice could not correct. Quantum loop gases, such as the toric code \cite{kitaev2003,fowlersurface} would likewise require much more complex constructions to be passively stabilized. Though beyond the scope of this work, a more general method for engineering shadow lattices to correct extended and multi-body errors in photonic systems would be an invaluable tool for future experiments in quantum simulation.

\section{Acknowledgments}

We would like to thank Jacob Taylor, John Chalker and Michael Foss-Feig for useful discussions. This material is based on work supported by EPSRC Grant Nos. EP/I032487/1 and EP/I031014/1, the University of Oxford, ARO MURI award (W911NF0910406) and NSF through the Physics Frontier Center at the Joint Quantum Institute.
\bibliography{SLbib,biblio}

\end{document}